\documentclass[letter,twocolumn]{jpsj3}
\usepackage{txfonts}
\usepackage[normalem]{ulem}
\usepackage{color}
\title{%
Magnetic Field Effect in One-Dimensional Charge Ordering Systems
}
\author{%
Yuichi \textsc{Otsuka}$^{1,}$\thanks{E-mail: otsukay@riken.jp}, 
Hitoshi \textsc{Seo}$^{2,3,4}$, and
Yukitoshi \textsc{Motome}$^{5}$
}
\inst{%
$^1$Advanced Institute for Computational Science, RIKEN, Kobe, Hyogo 650-0047, Japan\\
$^2$Condensed Matter Theory Laboratory, RIKEN, Wako, Saitama 351-0198, Japan\\
$^3$Quantum Matter Theory Research Team, RIKEN CEMS, Wako, Saitama 351-0198, Japan\\
$^4$JST, CREST, Wako, Saitama 351-0198, Japan\\
$^5$Department of Applied Physics, University of Tokyo, Bunkyo, Tokyo 113-8656, Japan
}
\recdate{\today}

\abst{%
We study the effects of an external magnetic field on charge
ordering in the one-dimensional extended Hubbard model at quarter
filling by the quantum Monte Carlo method. 
We find that the Zeeman coupling enhances the charge order correlation,
which is prominent when the system is located in the critical regime
near the charge ordering transition at zero magnetic field. 
This behavior is interpreted in terms of the crossover to the fully
spin-polarized limit where the model is exactly solvable.
Furthermore, by incorporating the interchain Coulomb repulsion, 
we show that the charge-ordering transition temperature is increased by
the magnetic field.
We also discuss the relevance of our results to magnetoresistance
effects observed in molecular conductors.
}

\begin{document}
\maketitle


In strongly correlated electron systems, the application of an external field
can lead to nontrivial responses owing to their collective nature.  
As a typical example, in molecular conductors, 
nonlinear electron conduction is widely observed in charge ordering (CO)
systems under an electric field~\cite{Kumai_Science1999,Sawano_Nature2005}.
On the other hand, the effects of a magnetic field in such CO systems
have been less intensively studied, 
although they have been widely investigated in
the case of metallic systems~\cite{Lebed_Book}.
Recently, it has been found that several molecular conductors, 
which are considered as potential CO systems,
exhibit relatively large positive magnetoresistance (MR) 
effects~\cite{Hanasaki_JPSJ2006,Yamaguchi_PRB2010}.
These phenomena were considered to be induced by the Zeeman effect, as
judged from their little dependence on the orientation of the applied magnetic
field.
This is in sharp contrast to the orbital effect, which generally plays
an important role in metallic molecular crystals~\cite{Lebed_Book}; 
owing to their anisotropic electronic structures, 
the responses are very sensitive to the orientation.

One such compound is 
TPP[Co(Pc)(CN)$_2$]$_2$~\cite{Hasegawa_JMC1998},
a highly one-dimensional (1D) system in which uniform columns
of Co(Pc)(CN)$_2$ molecules form a quarter-filled $\pi$-electronic band.
Its resistivity is semiconducting below about 250~K~\cite{Hasegawa_JMC1998},
and asymmetric NQR spectra are observed below 5~K,
suggesting the existence of CO correlation~\cite{Hanasaki_JPSJ2006},
although whether a long-range order exists in the ground state remains
elusive. 
The positive MR is prominent at low temperatures ($T$) of
$T \lesssim 20$~K~\cite{Hanasaki_JPSJ2006}. 
At $T=1.7$~K, for instance, 
there is no dependence on the direction of the applied magnetic field of
$B<10$~T, indicating the Zeeman effect as its origin~\cite{Hanasaki_JPSJ2006}. 
Another example is
$\theta$-(BEDT-TTF)$_2$CoZn(SCN)$_2$~\cite{Mori_PRB1998}.
This compound is a quasi-two-dimensional material
in which short-range CO correlations develop~\cite{Takahashi_JPSJ2006}; 
the observed positive MR at low $T$ is interpreted in terms of the spin
polarization in localized paramagnetic spins near the CO domain
boundaries, namely, the Zeeman effect, which suppresses the electron
conduction because of the Pauli exclusion principle~\cite{Yamaguchi_PRB2010}. 

In the CO insulating states, the spin degree of freedom typically
behaves as localized spin systems, analogous to the case of Mott
insulators.
This is a common feature of insulating states driven by strong
correlation. 
The remaining magnetic sector, therefore, can respond to a magnetic
field through the Zeeman coupling.
Theoretically, to describe such CO insulating states in molecular
conductors, the extended Hubbard model (EHM),
including not only the on-site but also the intersite Coulomb
interactions, has been widely used~\cite{Seo_JPSJ2006}.
However, little is known about the effect of the magnetic field in
the EHM even on the simplest lattice structures, which is the subject of
this study.

In this work, we numerically investigate the 1D quarter-filled EHM 
coupled to the magnetic field via the Zeeman term.
Although such a model was previously studied 
from several viewpoints~\cite{Kishigi_JPSJ2000,Kishigi_JPSJ2000a,Kitazawa_JPCM2003,Mancini_EPJB2008},
how CO is affected by the magnetic field has been less discussed.
By a quantum Monte Carlo method, we show that the CO 
correlation is enhanced by the applied magnetic field, and the behavior
is continuously connected to the fully spin-polarized limit where exact
results are known. 
We further take into account the interchain Coulomb interaction 
and show that the CO transition temperature is increased by the magnetic
field.


We begin with the EHM for a 1D chain in an applied magnetic field, 
whose Hamiltonian is given by,
 \begin{align}
  \mathcal{H} =&
  -t \sum_{i, \sigma}
  \left( 
  c_{i+1 \sigma}^{\dagger}c_{i \sigma} + \text{h.c.}
  \right)
    + U \sum_{i} n_{i  \uparrow} \ n_{i \downarrow} \nonumber\\
  & + V \sum_{i} n_{i  } \ n_{i  +1   }
  - h \sum_{i} \left( n_{i \uparrow} - n_{i \downarrow} \right),
  \label{eq:model}
 \end{align}
where 
$c_{i \sigma}$ ($c_{i \sigma}^{\dagger}$) 
is an annihilation (creation) operator of
an electron with spin $\sigma$ at site $i$, and
$n_{i \sigma} = c_{i \sigma}^{\dagger} c_{i \sigma}$
is the number operator ($n_{i} =n_{i \uparrow} + n_{i \downarrow}$).
The on-site and intersite Coulomb interactions are 
denoted by $U$ and $V$, respectively.
The last term represents the Zeeman coupling to the magnetic field $h$
applied in the $z$ direction.
We consider the case of quarter filling, i.e., 
$\sum_i \langle n_i \rangle/N = 1/2$ with $N$ being the number of sites.

The properties of this model at $h=0$ are well known~\cite{Seo_JPSJ2006}.
For repulsive interactions, the CO insulating phase appears at $T=0$ 
in the region of large ($U/t$, $V/t$) beyond a critical line;
otherwise, the system is metallic (Tomonaga-Luttinger liquid),
and the spin degree of freedom behaves as a 1D $S$=1/2 Heisenberg
model as a consequence of spin-charge separation~\cite{Yoshioka_JPSJ2000}.
The on-site Coulomb repulsion is fixed at $U/t=6$ in this paper,
where the CO phase transition occurs at a critical point of
$V_{\text{c}}/t \simeq 3.5$ for $h=0$~\cite{Ejima_EPL2005}.
At finite $T$, the long-range order is prevented by thermal fluctuations
in 1D systems. 

We investigate the 1D model by the quantum Monte Carlo
method based on the stochastic series expansion (SSE) 
with the operator-loop update~\cite{Sandvik_PRB1991,Sandvik_PRB1999,Sengupta_PRB2002}. 
An advantage of this method is that 
it can incorporate the magnetic field
unlike other methods that work on the canonical ensemble
with a fixed total magnetization.
The calculations fully include thermal and quantum fluctuations
in an unbiased manner.
We have studied the system consisting of $N=64$ sites under the periodic
boundary condition and confirmed that finite-size effects are negligible
down to $T/t=0.02$.
We calculate the charge and spin structure factors,
\begin{align}
 C(q)  &=  
 \frac{1}{N} \sum_{j k} \, e^{ i q (j-k) } \,
 \langle  \, 
 ( n_{j} - 1/2 )  \,
 ( n_{k} - 1/2 ) 
 \, \rangle \, ,
  \label{eq:charge-structure}\\
 S(q)  &=  
 \frac{1}{N} \sum_{j k} \, e^{ i q (j-k) } \,
 \langle  \, 
 ( n_{j \uparrow} - n_{j \downarrow}) \, 
 ( n_{k \uparrow} - n_{k \downarrow})  
 \, \rangle \, ,
  \label{eq:spin-structure}
\end{align}
respectively.

\begin{figure}
 \centering
 \includegraphics[width=0.45\textwidth]{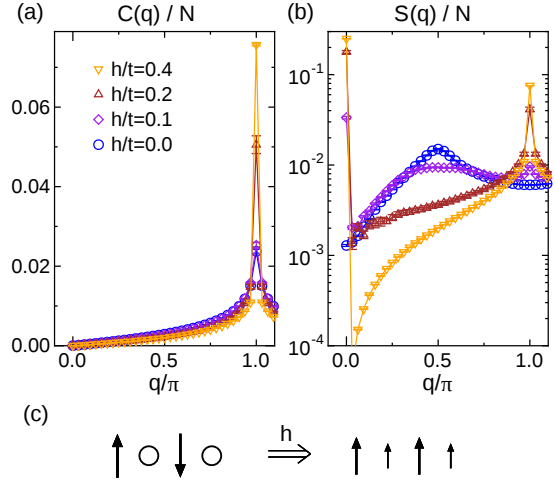}
 \caption{%
 (Color online) 
 Charge (a) and spin (b) structure factors of 
 the one-dimensional extended Hubbard model for $U/t=6$ and $V/t=3$
 under various values of the magnetic field $h/t$  at $T/t=0.05$ ($N=64$).
 Real-space patterns of the charge and
 spin structures under weak (left) and strong (right) magnetic fields 
 are schematically shown in (c).
 }\label{fig:structure} 
\end{figure}

First, let us discuss how the magnetic field affects the overall
features in the charge and spin degrees of freedom. 
In Fig.~\ref{fig:structure},
we show the charge and spin structure factors at $V/t=3$ and $T/t=0.05$ 
for various values of the magnetic field $h$.
In the absence of the magnetic field, the characteristic peaks of the
charge and spin structures are located at $q=\pi$ and $q=\pi/2$,
respectively.
The former represents the $4k_{\text{F}}$
($k_{\text{F}}$: Fermi momentum at $h=0$) 
CO correlation originating in the intersite interaction $V$.
Although the CO does not achieve a long-range order in the ground state
for this value of $V/t$, the CO correlation is dominant 
as the system is in the critical regime near $V_{\text{c}}$.
The peak of $S(q)$ at $q=\pi/2$ indicates an antiferromagnetic (AF)
correlation resulting from a superexchange coupling between every other
sites appearing in the fourth-order perturbation from the strong
coupling limit under CO.
Once the magnetic field is switched on, 
the CO peak $C(\pi)$ is notably enhanced; meanwhile, in the spin
sector, $S(\pi/2)$ becomes gradually blurred and two peaks 
at other momenta develop instead.
One is the ferromagnetic component $S(0)$, 
as a consequence of the magnetization induced by the applied magnetic
field. 
The other peak develops at $q=\pi$ along with the enhancement of CO; 
in fact, $S(\pi)$ approaches $C(\pi)$ in the fully
spin-polarized limit as discussed below.
The overall variation driven by the magnetic field is schematically
depicted in Fig.~\ref{fig:structure}(c). 

\begin{figure}
 \centering
 \includegraphics[width=0.40\textwidth]{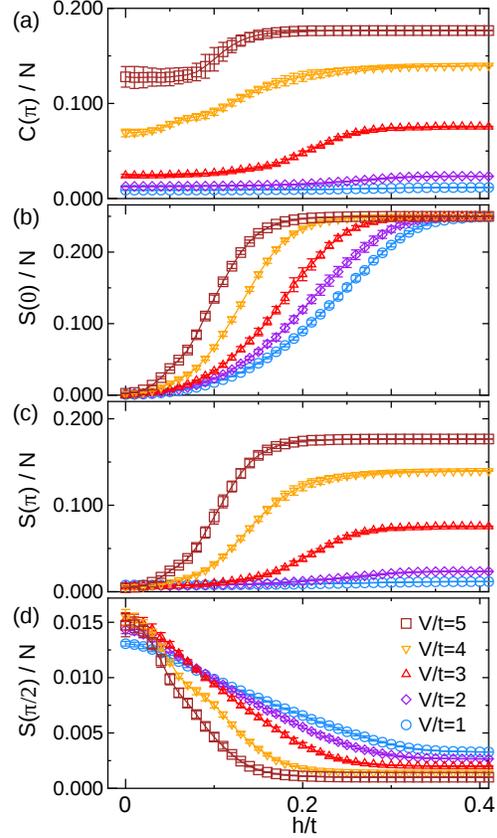}
 \caption{%
 (Color online) 
 Charge structure factor at $q=\pi$ (a) and spin structure factors 
 at $q=0$ (b), $q=\pi$ (c), and $q=\pi/2$ (d)  
 as functions of the magnetic field $h/t$  for $U/t=6$ and $T/t=0.05$
 ($N=64$).
 }\label{fig:h-dep}
\end{figure}

The magnetic field dependence of the characteristic peak values is
plotted in Fig.~\ref{fig:h-dep} for different values of $V/t$. 
All the data monotonically vary as $h/t$ increases.
The CO correlation $C(\pi)$ [Fig.~\ref{fig:h-dep}(a)]
develops together with the ferromagnetic component $S(0)$
[Fig.~\ref{fig:h-dep}(b)].
However, a large enhancement in the former is only observed for $V/t>2$,
which we interpret as follows.
The $h/t=\infty$ limit corresponds to the fully spin-polarized state, 
which is confirmed by $S(0)$ approaching the saturated value for large
$h/t$. 
In this limit, the system can be described by the 1D interacting spinless
fermion model~\cite{Schulz_PRL1990}:
\begin{align}
\mathcal{H}_{\rm SF} = 
 -t \sum_{i} \left( a_{i+1}^{\dagger}a_{i} + \text{h.c.}\right)
 + V \sum_{i} \tilde{n}_{i  } \ \tilde{n}_{i  +1},
\end{align}
where $a_{i}$ $(a_{i}^{\dagger})$ annihilates (creates) 
a spinless fermion at site $i$ and 
$\tilde{n}_{i}=a_{i}^{\dagger}a_{i}$ is the number operator.
Because of the perfect spin polarization, the system effectively becomes
half-filled, and the on-site interaction $U$ can be omitted.
The exact solution of this model gives the transition point,
$V_\text{c}/t=2$, between the metallic and CO insulating
states~\cite{Ovchinnikov_JETP1973,Haldane_PRL1980}.
Our result that the CO is enhanced mainly for $V/t>2$ is consistent
with this critical value of $V_\text{c}/t$,
since the system with $V/t < V_{\text{c}}/t$ remains metallic even in the
spin-polarized limit.

The crossover to the fully spin-polarized limit
is also confirmed in the behavior of other peak values.
The AF correlation $S(\pi/2)$, which is characteristic of the EHM at $h=0$,
is gradually depressed by the magnetic field, as shown in Fig.~\ref{fig:h-dep}(d).
This indicates that the virtual superexchange processes
become forbidden in the spin-polarized limit.
Furthermore, $C(q)$ and $S(q)$ defined in
Eqs.~(\ref{eq:charge-structure}) and (\ref{eq:spin-structure})
are identical to each other in this limit except for $q=0$
[for $q=\pi$, see Figs.~\ref{fig:h-dep}(a) and \ref{fig:h-dep}(c)].
This accounts for the above-mentioned development of $S(\pi)$
in Fig.~\ref{fig:structure}(b).

\begin{figure}
 \centering
 \includegraphics[width=0.40\textwidth]{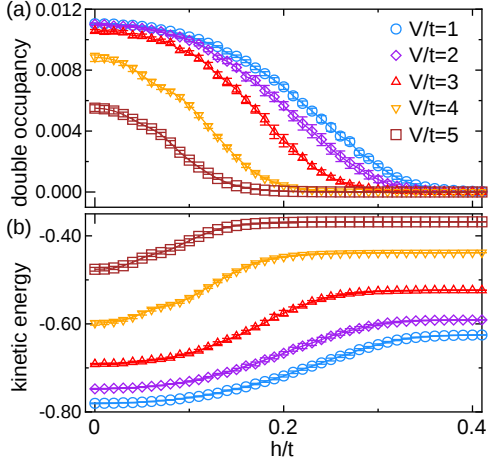} 
 \caption{%
 (Color online) 
 Double occupancy (a) and kinetic energy (b) at different values of $V/t$
 as functions of applied magnetic field $h/t$ for $U/t=6$ and $T/t=0.05$
 ($N=64$).
 }\label{fig:dk}
\end{figure}

The $h/t$-dependences of the double occupancy and kinetic energy are
shown in Fig.~\ref{fig:dk}.
When the spins become polarized by increasing $h/t$, 
the double occupancy is suppressed owing to the Pauli exclusion principle.
In addition, the effective filling is increased from a quarter to a half.
As a result, the kinetic-energy gain is also suppressed 
[Fig.~\ref{fig:dk}(b)].
These results indicate that the electrons approaching the spin-polarized
limit possess a more localized character than those of the EHM even at the
same value of $V/t$; 
the magnetic field effectively enhances the effect of $U/t$.
This is the reason why CO is stabilized by $h$. 

\begin{figure}
 \centering
 \includegraphics[width=0.40\textwidth]{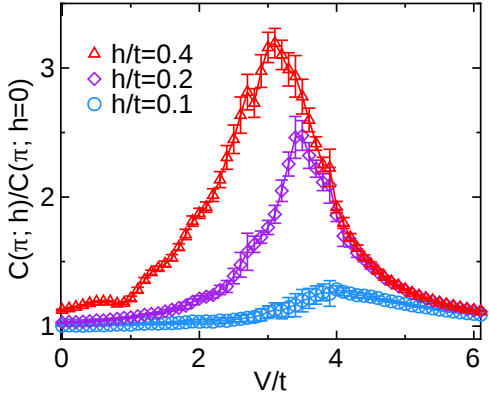}
 \caption{%
 (Color online) 
 Enhancement ratio of charge-order correlation $C(\pi)$
 between $h>0$ and $h=0$ as a function of $V/t$  for $U/t=6$ and
 $T/t=0.05$ ($N=64$).
 }
 \label{fig:V-dep}
\end{figure}

Another feature in Fig.~\ref{fig:h-dep}(a) is that the relative
enhancement in $C(\pi)$ is prominent at intermediate $V/t$.
This can be clearly observed from the enhancement ratio, as shown 
in Fig.~\ref{fig:V-dep}.
Here, the ratio of $C(\pi)$ for $h>0$ to that at $h=0$
is plotted as functions of $V/t$.
We find that the peaks are located at $V/t$=3 -- 4, 
which is in the vicinity of the critical point 
$V_{\text{c}}/t \simeq 3.5$ at $h/t=0$.
Furthermore, the value of $V/t$ at which the ratio is maximum
becomes smaller when the magnetic field is increased.
This indicates that the CO region in the phase diagram
expands to a smaller $U$-$V$ region.

The enhancement of CO by the magnetic field is not
directly anticipated from the studies for $h=0$.
For instance, the bosonization studies for $h=0$ have clarified that 
the low-energy properties of the EHM are described by 
the phase Hamiltonian with spin-charge separation,
and the CO instability originates in 
the higher-order Umklapp scattering in the charge
part~\cite{Yoshioka_JPSJ2000}.
However, our observation that the magnetic field enhances CO 
implies the strong interplay between the spin and charge degrees of
freedom under a magnetic field~\cite{Yoshioka_JKPS2013}.

Next, we consider the effect of the magnetic field on the finite-$T$ CO
phase transition by including the interchain coupling term to the model 
in Eq.~(\ref{eq:model}). 
We treat the interchain Coulomb repulsion between neighboring chains
$V_{\perp}$ within the mean-field approximation, as in our previous
works~\cite{Seo-Motome_JPSJ2007,Otsuka_JPSJ2008,Yoshioka_Crystals2012}.
The resulting additional term to Eq.~(\ref{eq:model}) is 
\begin{equation}
 \mathcal{H}_{\perp}^{\text{MF}} = 
  z V_{\perp} n_{\text{CO}} \sum_{i} (-1)^{i} n_{i}
  + \frac{1}{2} z N V_{\perp} n_{\text{CO}}^{2},
  \label{eq:vperp}
\end{equation}
where $n_{\text{CO}}$ is the CO order parameter with twofold
periodicity determined self-consistently
and $z$ is the number of nearest-neighbor chains, which we take as 2 
in this work.
This interchain term gives rise to the finite-$T$ CO phase transition.

\begin{figure}
 \centering
 \includegraphics[width=0.40\textwidth]{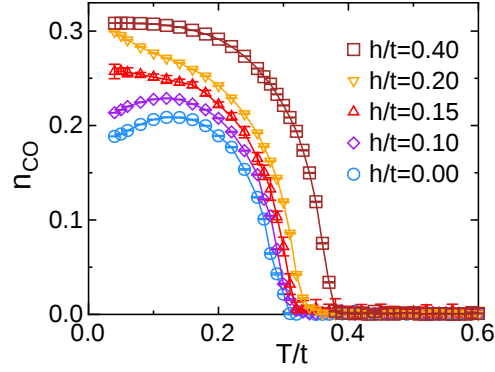}
 \caption{%
 (Color online) 
 Temperature dependence of the charge-ordering order parameter 
 for $U/t=6$, $V/t=2.5$, and $V_{\perp}/t=0.5$ ($N=64$).
 }\label{fig:ICMF}
\end{figure}

\begin{figure}
 \centering
 \includegraphics[width=0.40\textwidth]{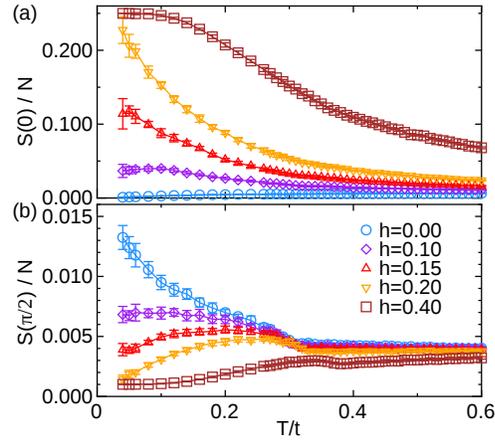}
 \caption{%
 (Color online) 
 Temperature dependence of 
 spin structure factors at $q=0$ (a) and $q=\pi/2$ (b)  
 for $U/t=6$, $V/t=2.5$, and $V_{\perp}/t=0.5$ ($N=64$).
 }\label{fig:ICMF_spin}
\end{figure}

The results show that the transition temperature $T_{\text{CO}}$ is
increased by applying the magnetic field, as expected from the 
results that we have observed for the 1D model.
The $T$ dependence of the CO order parameter $n_{\text{CO}}$ is plotted
in Fig.~\ref{fig:ICMF}.
For $V/t=2.5$ and $V_{\perp}/t=0.5$, $T_{\text{CO}}/t$
is increased monotonically 
from $\sim 0.3$ at $h/t=0$ to $\sim 0.4$ at $h/t=0.4$,
where the system nearly reaches the spin-polarized regime.
We note that $n_{\text{CO}}$ slightly decreases 
at low $T$ for $h/t \lesssim 0.1$, which was ascribed to 
the development of the $2k_{\text{F}}$-AF correlations 
caused by the superexchange interaction between the charge-rich
sites in the case of $h=0$~\cite{Otsuka_JPSJ2008}.
On the other hand, in higher magnetic fields,
the dominant magnetic correlation becomes ferromagnetic;
$S(0)$ develops and thereby $S(\pi/2)$ is suppressed at low $T$,
as shown in Fig.~\ref{fig:ICMF_spin}.
Consequently, CO becomes further stabilized
for $h/t \gtrsim 0.15$ with the decrease in $T/t$.

Now, let us discuss our results in comparison with experiments.
Our finding that the magnetic field stabilizes CO 
is consistent with
the positive MR effect observed in quasi-1D molecular conductors 
such as TPP[Co(Pc)(CN)$_2$]$_2$~\cite{Hasegawa_JMC1998}.
In the 1D systems,
the Umklapp scattering is responsible for
increasing the resistivity as well as stabilizing
CO at $h=0$~\cite{Yoshioka_JPSJ2000}.
Thus, the positive MR can be interpreted as a consequence of 
the enhancement of such Umklapp processes under a magnetic field,
which is detected as the stabilized CO in our calculations.
On the other hand, the same explanation is not directly applicable to
the MR effect observed in
$\theta$-(BEDT-TTF)$_2$CoZn(SCN)$_2$~\cite{Yamaguchi_PRB2010},
even though it is also positive.
The $\theta$-(BEDT-TTF)$_2X$ family has a two-dimensional (2D) band
structure, 
and furthermore, the CO phenomena in these compounds are considered to
involve additional ingredients such as geometrical frustration and
coupling to the lattice degree of
freedom~\cite{Merino_PRB2005,Seo_JPSJ2006,Udagawa-Motome_PRL2007}. 
Such situations are beyond our simple 1D model.
Indeed, recent measurements on related 2D compounds, where the CO
transition is more clearly seen than in the above-mentioned materials,
indicate that the MR ranges from positive to negative near $T_{\text{CO}}$
depending on the compound~\cite{TakahashiMori_Communications}.
Although it may be difficult to discuss such experiments,
we expect that CO is a key ingredient for the MR effects.

Finally, we discuss our results in relation to the giant negative MR
effect observed in TPP[Fe(Pc)(CN)$_2$]$_2$. 
This material is an analog of the Pc-based compound mentioned above; 
by substituting Co with Fe, the MR
changes from positive to negative 
(and large)~\cite{Hanasaki_JPSJ2006b},
and its origin has been discussed
theoretically~\cite{Hotta_PRL2005,Hotta_PRB2010}.
In this compound, localized magnetic moments from $d$ electrons of Fe atoms in
the center of each molecule have strong Ising anisotropy, and an AF
correlation between them is observed~\cite{Tajima_PRB2008}.
As a result, $\pi$ electrons on the Pc unit, which form the 1D
conduction band, are subjected to the effective staggered magnetic field
from the $d$ electrons through $\pi$-$d$ coupling.
In a previous work~\cite{Otsuka_PhysicaB2010}, 
the authors found that the CO correlation is
also monotonically enhanced by the staggered magnetic field.
Hence, a uniform magnetic field brings about two
opposite effects on CO in the actual material.
Namely, although it enhances the CO correlation in $\pi$ electrons,
as shown in the current study, it will disturb the AF correlation between
the $d$ spins and suppress the effect that enhances CO through the
$\pi$-$d$ coupling.
It might depend on the parameters, such as the $\pi$-$d$ coupling and 
$g$-factors for the $\pi$ and $d$ electrons, whether CO is eventually
enhanced or depressed by the applied magnetic field.
The experiments showing the negative MR suggest that 
the negative effect through the $\pi$-$d$ coupling is larger in
TPP[Fe(Pc)(CN)$_2$]$_2$.
Although the competition was studied in terms of the charge
gap~\cite{Hotta_PRB2010},
it will be helpful to perform comprehensive studies including finite-$T$
effects on the MR phenomena.
This is left for a future study.

In summary, 
we have investigated the effects of the applied magnetic field on the
one-dimensional extended Hubbard model at quarter filling 
by the quantum Monte Carlo method.
The charge-order correlation is monotonically enhanced by the magnetic field.
The behavior is understood on the basis of a crossover to the fully
spin-polarized limit.
The enhancement of charge order is prominent near the phase boundary
between the paramagnetic metal and the charge-ordered insulator
in the absence of the magnetic field.
We have shown that the magnetic field increases the transition
temperature to the charge-ordered state within an interchain mean-field
approach.

\section*{Acknowledgments}
The authors would like to thank 
H.~Yoshioka, M.~Tsuchiizu,
H.~Mori, K.~Takahashi, and N.~Hanasaki
for fruitful discussions. 
This work was supported by 
Grants-in-Aid for Scientific Research (Nos. 24108511 and 24340076) from MEXT,
the Strategic Programs for Innovative Research (SPIRE), MEXT, 
the Computational Materials Science Initiative (CMSI), Japan,
and the RIKEN iTHES Project.


\end{document}